%% file: trombetta_proceedingLHCP2017_ArXiv.tex
\documentclass[10pt]{article}
\usepackage{graphicx}

\def\Title#1{\begin{center} {\Large #1 } \end{center}}
\def\Author#1{\begin{center}{ \sc #1} \end{center}}
\def\Address#1{\begin{center}{ \it #1} \end{center}}

\newcommand\pubblock{\rightline{\begin{tabular}{l} Proceedings of the Fifth Annual LHCP\\ \pubnumber\\
         \pubdate  \end{tabular}}}

\newenvironment{Abstract}{\begin{quotation} \begin{center} 
             \large ABSTRACT \end{center}\bigskip 
      \begin{center}\begin{large}}{\end{large}\end{center} \end{quotation}}

\newenvironment{Presented}{\begin{quotation} \begin{center} 
             PRESENTED AT\end{center}\bigskip 
      \begin{center}\begin{large}}{\end{large}\end{center} \end{quotation}}

\input econfmacros.tex
\usepackage{color}

 \newcommand\pubnumber{ }

\newcommand\pubdate{\today}

\def\affiliation{
On behalf of the ALICE Collaboration, \\
Dipartimento Interateneo di Fisica 'M.Merlin' \\
and Sezione INFN, Bari, Italy}

\begin{document}

% large size for the first page
\large
\begin{titlepage}
\pubblock

%% Change the title, name, abstract
%% Title 
\vfill
\Title{  NEW RESULTS ON INITIAL STATE AND QUARKONIA WITH ALICE  }
\vfill

%  if you need to add the support use this, fill the \support definition above. 
%   \Author{ FIRSTNAME LASTNAME \support }
\Author{ GIUSEPPE TROMBETTA  }
\Address{\affiliation}
\vfill
\begin{Abstract}

% \linenumbers

The study of quarkonia in heavy-ion collisions
has been the subject of intense experimental and theoretical effort, ever since their production was predicted
to be sensitive to the formation of a deconfined state of strongly-interacting matter, known as the Quark--Gluon Plasma (QGP).
In p--Pb collisions, Cold Nuclear Matter (CNM) effects, such as nuclear shadowing or partonic energy loss, are expected to influence quarkonium production. 
% Such effects are expected to play a role on top of ``hot'' effects such as QGP in Pb--Pb collisions. 
The study of such system is therefore crucial to shed light on the mechanisms taking place at the initial-state of quarkonium production,
and to disentangle the cold and hot nuclear effects envisioned in Pb--Pb collisions.
The ALICE experiment at the LHC, is capable of reconstructing J/$\psi$, $\psi$(2S) and $\Upsilon$ states at forward rapidity through 
their $\mu^{\rm{+}}\mu^{\rm{-}}$ decay channel, as well as J/$\psi$ at central rapidity through their $e^{\rm{+}}e^{\rm{-}}$ decay channel, 
down to zero transverse momentum. A review of the main ALICE findings from the measurements of the inclusive quarkonium yields 
in p--Pb collisions at $\sqrt{s_{\rm{NN}}}$ = 5.02 TeV, collected during the LHC Run I period, as well as more recent results from 
J/$\psi$ measurements in p--Pb at $\sqrt{s_{\rm{NN}}}$ = 8.16 TeV, from LHC Run II period, will be presented in this paper.

\end{Abstract}
\vfill

% DO NOT CHANGE 
\begin{Presented}
The Fifth Annual Conference\\
 on Large Hadron Collider Physics \\
Shanghai Jiao Tong University, Shanghai, China\\ 
May 15-20, 2017
\end{Presented}
\vfill
\end{titlepage}
\def\thefootnote{\fnsymbol{footnote}}
\setcounter{footnote}{0}
%

% normal size for the rest
\normalsize 

%% Your paper should be entered below. 

% \linenumbers

\section{Introduction}

The production of quarkonia -- bound states of heavy charm or beauty quark and anti-quark pairs -- represents a challenging and not yet fully understood subject in the field of High-Energy Physics, 
which invokes aspects of both the perturbative and non-perturbative regimes of Quantum Chromodynamics (QCD).
As of today, quarkonium formation in elementary hadronic systems is described by QCD-based models as a two-step 
process which involves the perturbative creation of the $q\bar{q}$ pair via hard scattering, and its subsequent non-perturbative
evolution towards a specific bound final state.
In heavy-ion systems, quarkonia play a further important role as they have long been proposed as ideal probes of the hot and dense 
strongly-interacting matter produced in such collisions. A suppression of charmonium production was indeed predicted as 
a signature of the phase transition of hadronic matter to a Quark--Gluon Plasma at sufficiently high energy densities \cite{Matsui:1986dk}.\\
Several effects due to the presence of Cold Nuclear Matter, and not related to the formation of QGP, can however contribute
to modify the observed quarkonium yields with respect to elementary nucleon--nucleon collisions.
During the initial stage of their formation, quarkonium production cross sections can be either suppressed (shadowing) 
or enhanced (anti-shadowing), as a result of the modification of the kinematical distributions of partons experienced in nuclei \cite{Eskola:2009uj}. 
Furthermore, if the production process is dominated by low-momentum gluons, as it is expected at LHC energies, the gluons may behave 
as a coherent and dense partonic system, which can be described in the framework of Color--Glass Condensate effective theory, 
providing predictions when combined with a specific quarkonium production model \cite{Ma:2015sia,Ducloue:2015gfa}. 
In addition, both the incoming partons and the outgoing $q\bar{q}$ pair propagating through the nucleus may lose energy by gluon radiation 
at various stages of the formation process, including the occurrence of coherent energy loss processes \cite{Arleo:2012rs}.
The evolving $q\bar{q}$ pair or even the final-state resonance may finally interact with the nucleons while traveling 
through the nucleus or with the other co-moving partons and hadrons produced in the collision \cite{Ferreiro:2014bia}, 
consequently losing energy or breaking up into open-flavor meson pairs.\\
The study of p--Pb collisions at the LHC provides an ideal way to test the interplay between the different 
mechanisms affecting quarkonium production in a previously unexplored kinematic range, where the yield is expected 
to be dominated by initial-state CNM effects.

\section{Quarkonium measurement with ALICE}

ALICE at the CERN LHC allows the measurement of quarkonium states down to zero transverse momentum ($p_{\rm T}$),
in a complementary kinematic region with respect to other LHC experiments \cite{Aamodt:2008zz}. 
At central rapidity ($y$), the J/$\psi$ resonance is detected through its $e^{\rm{+}}e^{\rm{-}}$ decay channel making use of
ALICE central barrel detectors, which cover the pseudorapidity range $|\eta| < 0.9$. At forward rapidity, J/$\psi$, $\psi$(2S) and $\Upsilon$ states 
are reconstructed through their $\mu^{\rm{+}}\mu^{\rm{-}}$ decay channel by means of the ALICE muon spectrometer, covering the pseudorapidity 
range $-4 < \eta < -2.5$.\\ 
The Time Projection Chamber (TPC) is the main tracking detector of the barrel, consisting of a large cylindrical drift 
chamber which allows also charged particle identification through specific energy loss (d$E$/d$x$) measurements. 
The Inner Tracking System (ITS) is a cylindrically-shaped tracker made up of six layers of silicon detectors which provide precise tracking and 
vertex reconstruction close to the interaction point. The muon spectrometer consists of a front absorber, used to filter out the hadrons 
produced in the interaction, a tracking system, made up of five Cathod Pad Chamber stations, a large 3 T$\cdot$m dipole magnet, and two 
Resistive Plate Chamber trigger stations, shielded by a muon filter wall. The information from the Zero Degree Calorimeters (ZDC), 
symmetrically placed at 112.5 m from the interaction point, is used to reject de-bunched proton-lead collisions, 
while two scintillator hodoscopes (V0), with pseudorapidity coverage  $2.8 < \eta < 5.1$ and $-3.7 < \eta < -1.7$, 
are used to remove beam-induced background.
A signal coincidence in the two V0 detectors provides the trigger to select Minimum Bias (MB) events on which
the dielectron analyses are performed, whereas dimuon analyses rely on a dimuon trigger which requires, in addition
to the MB condition, the detection of two opposite-sign tracks in the muon trigger stations.
Finally, the slow neutron energy deposited by the Pb nucleus remnants in the ZDC is used to determine the event centrality.
Such observable was found to be less sensitive to the dynamical bias observed in centrality estimations based on charged-particle 
multiplicity, which are usually employed in Pb--Pb collisions \cite{Adam:2014qja}.\\
Over the last years, ALICE collected p--Pb data in two beam configurations, corresponding to 
to either protons or lead ions going towards the muon spectrometer, which allowed the coverage of 
two different intervals in the forward and backward dimuon rapidity regions and of two corresponding 
Bjorken-x ($x_{\rm Bj}$) ranges. 
At the beginning of 2013, during the LHC Run I period, data samples from p--Pb collisions at the centre-of-mass energy 
per nucleon-nucleon collision $\sqrt{s_{\rm{NN}}}$ = 5.02 TeV were collected, with corresponding integrated luminosities 
of 5 nb$^{-1}$, 5.8 nb$^{-1}$ and 51 $\mu \rm b^{-1}$ for the forward, backward and central rapidity intervals. 
The analysis of such samples provided a wide variety of physics results which helped investigating 
the size of CNM effects on both charmonium and bottomonium production. 
In 2016, p--Pb collisions at $\sqrt{s_{\rm{NN}}}$ = 8.16 TeV delivered during the LHC Run II period allowed even
larger data samples to be collected, with corresponding integrated luminosities of 8.7 nb$^{-1}$ and 12.9 nb$^{-1}$  
in the forward and backward rapidity intervals, on which dimuon analyses of $J/\psi$ production have recently been carried out.

% Observation of the Higgs Boson,  \cite{Aad:2012tfa},\cite{Chatrchyan:2012ufa}. 

%%
%%   use this format to include an .eps figure into your paper
%%
% \begin{figure}[htb]
% \centering
% \includegraphics[height=2in]{head_lhcp2017.jpg}
% \caption{ Place the caption here}
% \label{fig:figure1}
% \end{figure}
% %%%%%%%%%%%%%%%%%%%%%%%%%%%%%%%%%%%%%%%%%%%%%%%%%%%%%%%%%%%%%%%%%%%%%%%%%%%
% 
% See Figure \ref{fig:figure1} and Table \ref{tab:table1}. 
%  
% %%%%%%%%%%%%%%%%%%%%%%%%%%%%%%%%%%%%%%%%%%%%%%%%%%%%%%%%%%%%%%%%%%%%%%%%%
% 
% %%%%%%%%%%%%%%%%%%%%%%%%%%%%%%%%%%%%%%%%%%%%%%%%%%%%%%%%%%%%%%%%%%%%%%%%%
% %%
% %%   use this format to include a LaTeX table  into your paper
% %%
% \begin{table}[t]
% \begin{center}
% \begin{tabular}{l|ccc}  
% Patient &  Initial level($\mu$g/cc) &  w. Magnet &  
% w. Magnet and Sound \\ \hline
%  Guglielmo B.  &   0.12     &     0.10      &     0.001  \\
%  Ferrando di N. &  0.15     &     0.11      &  $< 0.0005$ \\ \hline
% \end{tabular}
% \caption{ place the caption here }
% \label{tab:table1}
% \end{center}
% \end{table}
% %%%%%%%%%%%%%%%%%%%%%%%%%%%%%%%%%%%%%%%%%%%%%%%%%%%%%%%%%%%%%%%%%%%%%%%%%%%

\section{Results from Quarkonium analyses in p--Pb collisions}

The production of J/$\psi$ in p--Pb collisions at $\sqrt{s_{\rm{NN}}}$ = 5.02 TeV has been deeply studied by ALICE,
carrying out measurements of the J/$\psi$ yield either as a function of $y$ and $p_{\rm T}$ \cite{Abelev:2013yxa,Adam:2015iga}, 
or as a function of centrality \cite{Adam:2015jsa}. In order to quantify the size of nuclear effects on their production, 
the nuclear modification factor ($R_{\rm pPb}$), i.e. the ratio of the yield measured in p--Pb collisions to that in pp scaled 
by the number of binary collisions, was employed. The first measurements showed that the J/$\psi$ yield at forward rapidity 
was significantly suppressed while at backward rapidity the production was consistent with a binary scaling from pp 
collisions. Later, differential measurements as a function of transverse momentum, pointed out that the suppression 
was strongly dependent on $p_{\rm T}$, being significantly larger at forward $y$ and low $p_{\rm T}$ while tending
to vanish towards high $p_{\rm T}$ and at backward $y$. 
The results were found to be in fair agreement with theoretical calculations including nuclear 
shadowing and/or energy loss, as well as with CGC-inspired models.
% In Figure ??, the measurements of the J/$\psi$ nuclear modification factor as a function of the estimated collision
% centrality, discussed in ??, have been reported.  
Further measurements as a function of centrality, in three rapidity intervals, indicated that while data at backward
$y$ were consistent with no nuclear modifications for the most peripheral and semi-central events, a significant suppression of the
J/$\psi$ production in the full centrality range was present in the middle and forward rapidity intervals, whereas a
hint for enhancement was observed in the most central collisions at backward rapidity.

\begin{figure}[b]
\centering
\includegraphics[width=0.329\textwidth]{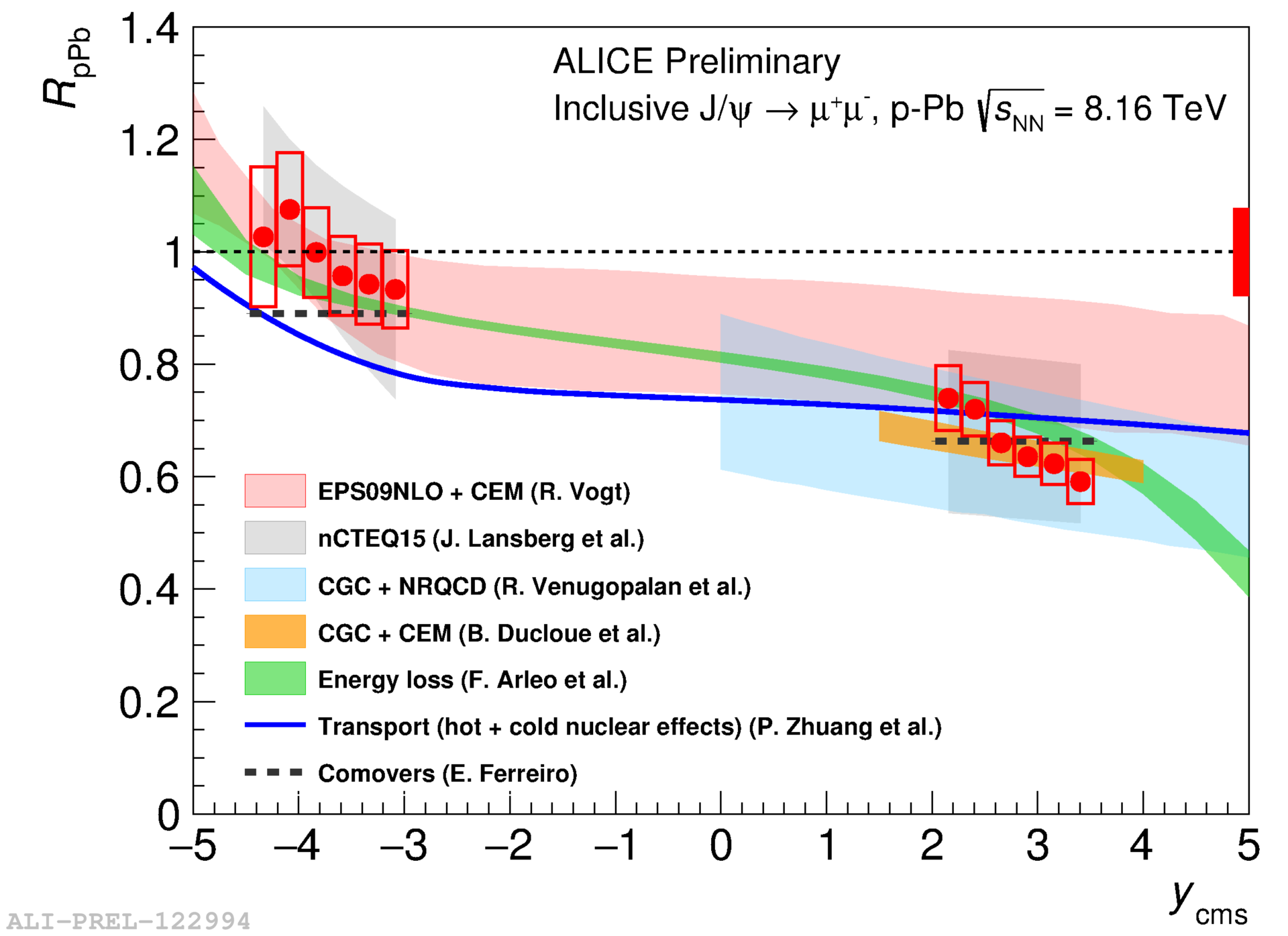}
\includegraphics[width=0.329\textwidth]{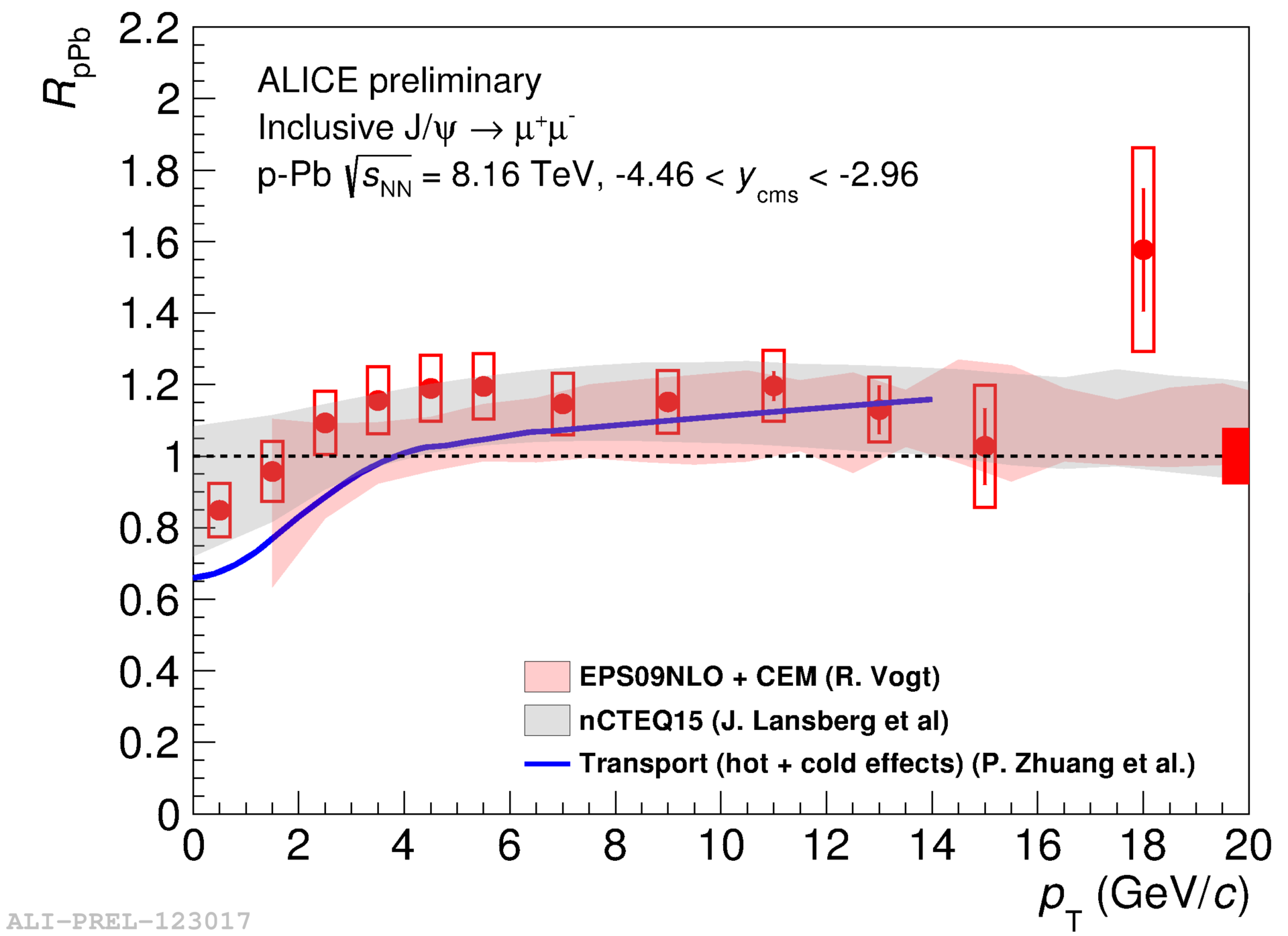}
\includegraphics[width=0.329\textwidth]{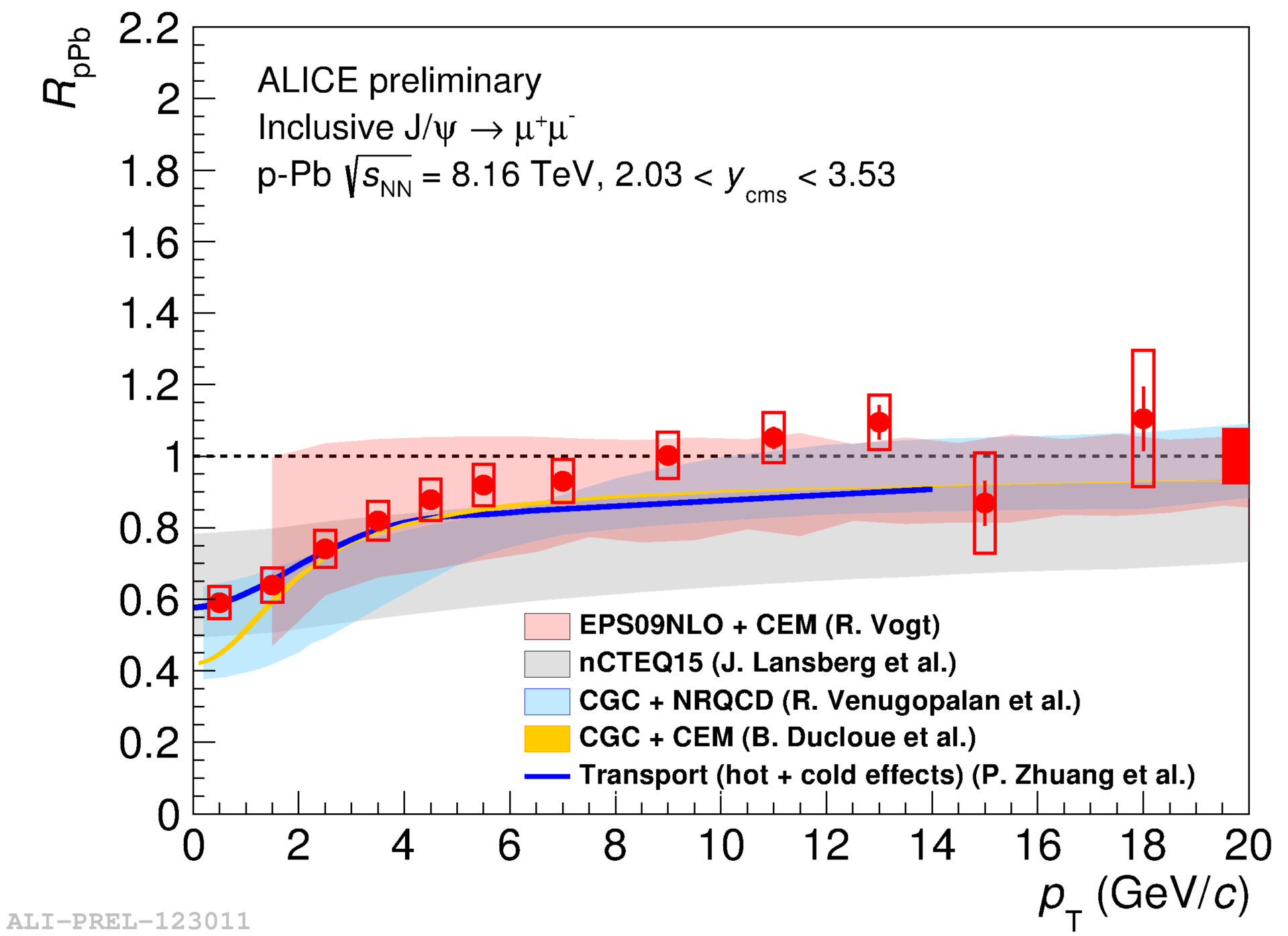}
\caption{The nuclear modification factor for inclusive J/$\psi$ production at $\sqrt{s_{\rm{NN}}}$ = 8.16 TeV as a function 
of rapidity (Left) and as a function of transverse momentum, at backward (Middle) and forward (Right) $y$, compared to theory 
predictions. The error bars represent statistical uncertainties, the boxes around the points the uncorrelated systematic 
uncertainties and the filled boxes around $R_{\rm pPb} = 1$ represent the correlated global uncertainties.}
\label{fig:figure1}
\end{figure}

\noindent The recent analysis of p--Pb data at $\sqrt{s_{\rm{NN}}}$ = 8.16 TeV, collected during the LHC Run II period, provided an
% extension in the probed $x_{\rm Bj}$ values for the nucleuons in the Pb nucleus with respect to previous J/$\psi$ studies 
% (from  $2\cdot10^{-5} < x_{\rm Bj} < 8\cdot10^{-5}$ and $1\cdot10^{-2} < x_{\rm Bj} < 5\cdot10^{-2}$ at $\sqrt{s_{\rm{NN}}}$ = 5.02 TeV,
% to $1\cdot10^{-5} < x_{\rm Bj} < 5\cdot10^{-5}$ and $7\cdot10^{-3} < x_{\rm Bj} < 3\cdot10^{-2}$ at $\sqrt{s_{\rm{NN}}}$ = 8.16 TeV, in the forward and backward rapidity
% regions respectively) as well as an 
even more detailed comparison to model predictions, thanks to the unprecedented reach in integrated luminosity.
In Figure \ref{fig:figure1}, ALICE measurements of the nuclear modification factor for inclusive J/$\psi$ production 
in p--Pb collisions at $\sqrt{s_{\rm{NN}}}$ = 8.16 TeV \cite{PAS:alice} are reported as a function of $y$, over the rapidity
intervals  $-4.46 < y < -2.96$ and $2.03 < y < 3.53$, and as a function of transverse momentum, 
up to $p_{\rm T} = 20$ GeV/$c$. While a significant suppression has been measured at forward rapidity, no significant 
nuclear effects are seen at backward rapidity, revealing a compatible trend to previous measurements at 
$\sqrt{s_{\rm{NN}}}$ = 5.02 TeV. Results as a function of $p_{\rm T}$ show hints for an enhancement of the production 
at $p_{\rm T} > 4$ GeV/$c$ at backward $y$, whereas at forward $y$ the $R_{\rm pPb}$ tends to unity towards high-$p_{\rm T}$.
Data are compared with various theoretical calculations implementing different combinations of CNM effects. In particular,
predictions from models including shadowing parametrizations \cite{Albacete:2013ei,Lansberg:2016deg}, CGC-based approaches \cite{Ma:2015sia,Ducloue:2015gfa} 
and coherent energy loss mechanisms \cite{Arleo:2012rs}, in combination with different production models, are reported. 
Central predictions from models based on a comover approach \cite{Ferreiro:2014bia} or including a combination of CNM effects and interactions 
with the produced medium \cite{Chen:2016dke} are also compared to data. All the calculations appear reasonably in agreement with the data,
although it should be noted that data achieved a precision which is challenging for most model predictions.\\ 

\begin{figure}[h]
\centering
\includegraphics[width=0.329\textwidth]{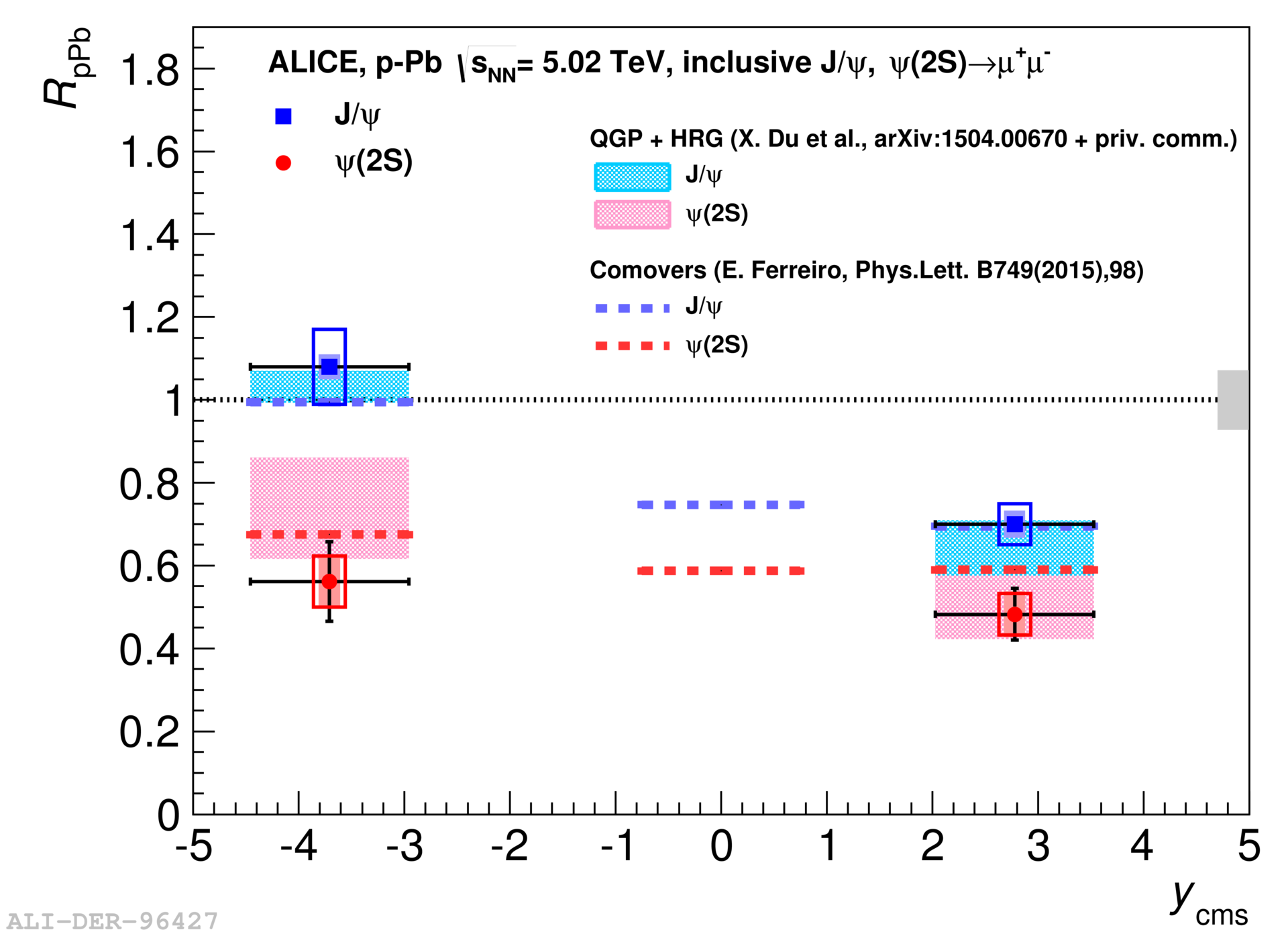}
\includegraphics[width=0.329\textwidth]{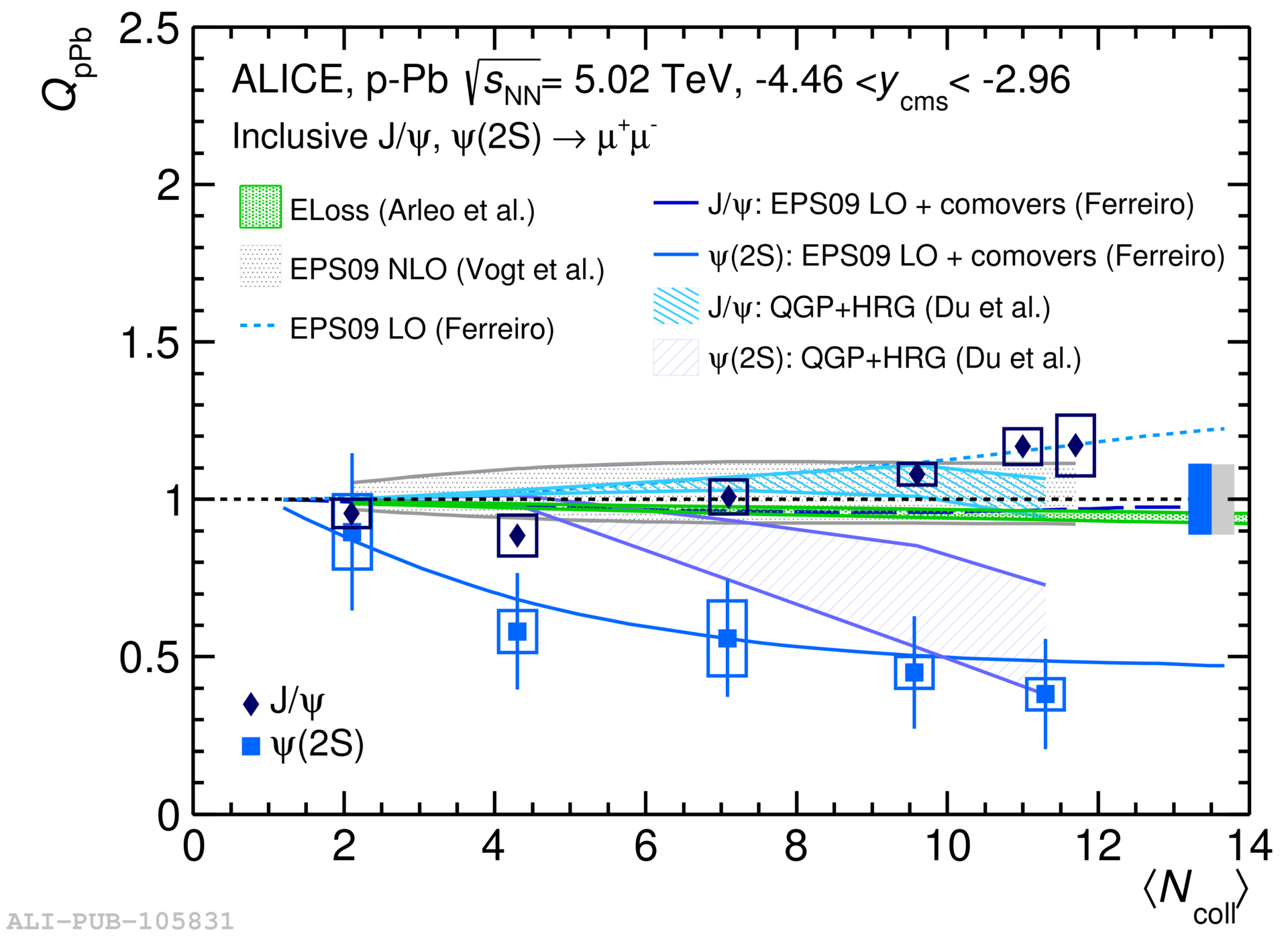}
\includegraphics[width=0.329\textwidth]{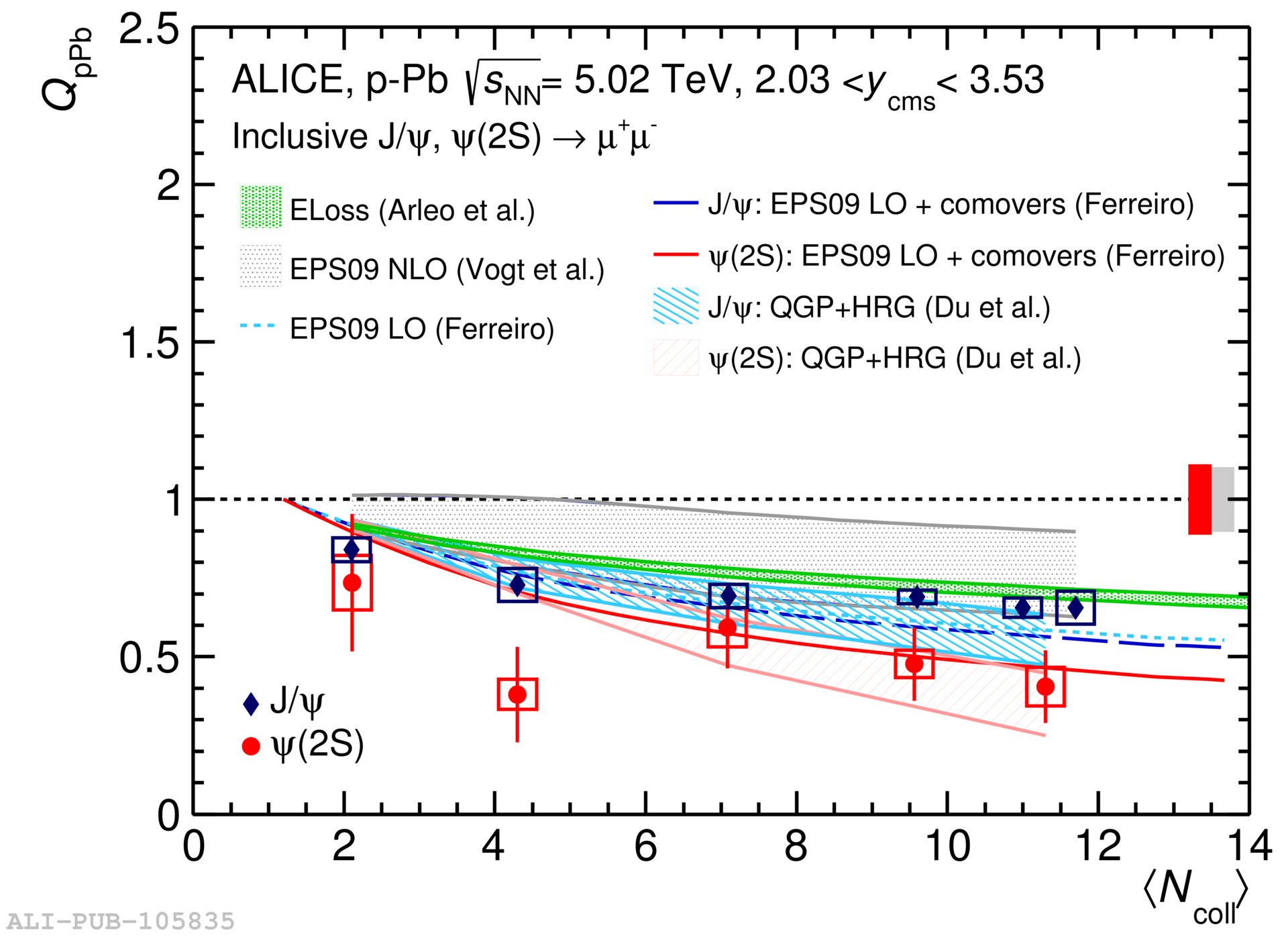}
\caption{The nuclear modification factor for inclusive $\psi \rm(2S)$ production in p--Pb collisions at $\sqrt{s_{\rm{NN}}}$ = 5.02 TeV 
as a function of rapidity (Left) and of centrality at backward (Middle) and forward (Right) $y$. Data are compared to similar measurements
for J/$\psi$, as well as to different theory predictions. The error bars represent statistical uncertainties, the boxes around the points the uncorrelated systematic 
uncertainties and the filled boxes around $R_{\rm pPb} = 1$ represent the correlated global uncertainties.}
\label{fig:figure2}
\end{figure}

\noindent The nuclear modification factor for the $\psi \rm(2S)$ state production was measured by ALICE in p--Pb collisions at 
$\sqrt{s_{\rm{NN}}}$ = 5.02 TeV, integrated over $p_{\rm T}$, as a function of rapidity \cite{Abelev:2014zpa} and centrality \cite{Adam:2015jsa},
and is reported in Figure \ref{fig:figure2}. Measurements showed a significantly stronger suppression of the $\psi \rm(2S)$ with 
respect to the J/$\psi$ in the same kinematic ranges, especially at backward $y$, which increases with the centrality of 
the collision. Considering that, within the accessed kinematic domains, the time needed to form the final-state resonance 
is larger than the crossing time of the $c\bar{c}$ pair in the nucleus, such observation cannot be ascribed to a break-up
by cold nuclear matter of the more weakly bound $\psi \rm(2S)$ state. Furthermore, CNM models based on shadowing and parton energy loss, 
being independent on quantum numbers of the final-state, predict the same degree of suppression for both charmonium states
and fail in reproducing $\psi \rm(2S)$ experimental results. Final-state effects need to be introduced to explain the different
suppression. Models including a dissociation contribution with co-moving partons and particles \cite{Ferreiro:2014bia}, or with a hot hadron resonance
gas experiencing a short phase transition to a QGP phase \cite{Zhao:2010nk}, appear indeed capable of reproducing the size of the observed suppression.\\

\begin{figure}[h]
\centering
\includegraphics[width=0.35\textwidth]{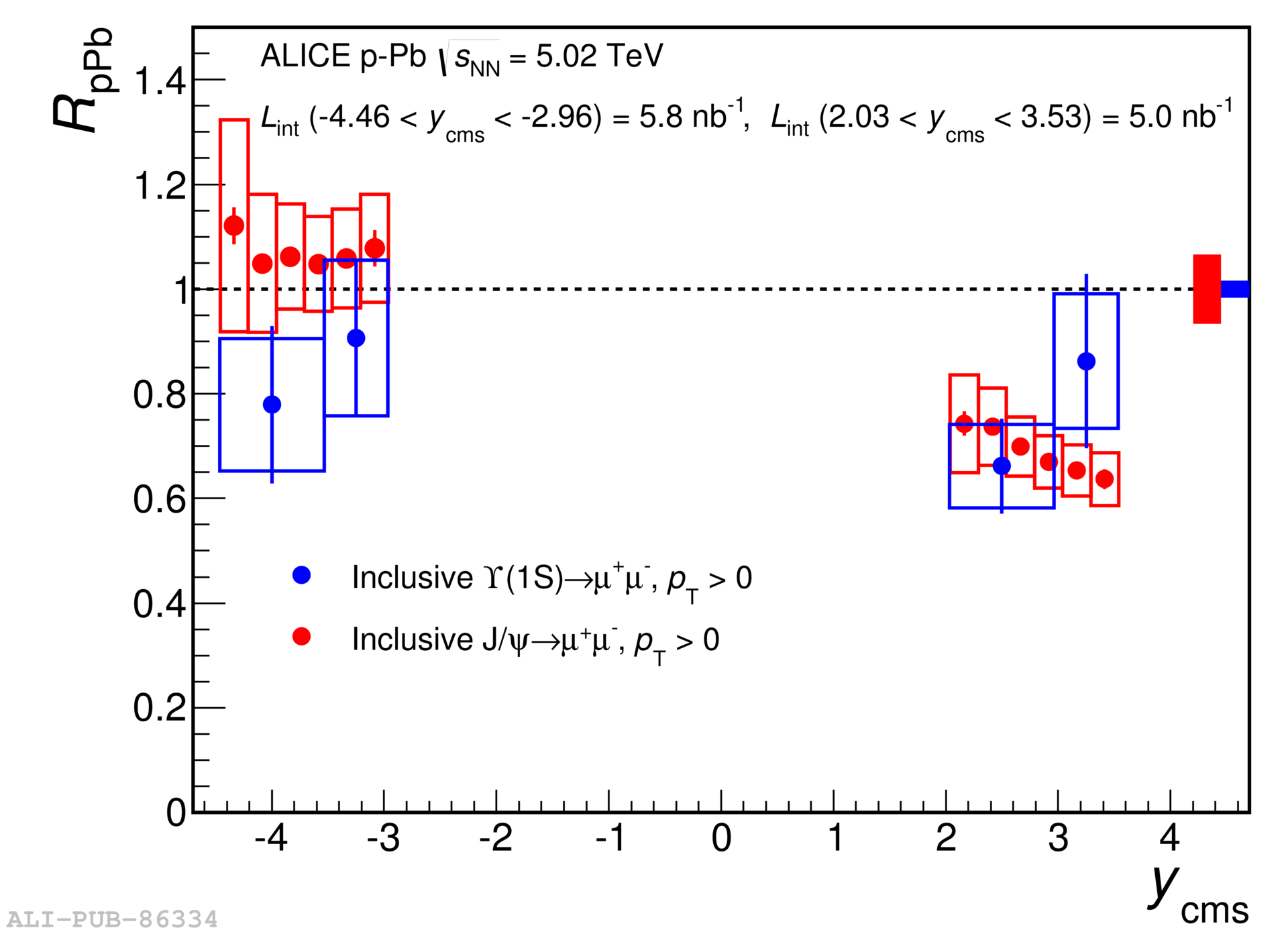}
\includegraphics[width=0.35\textwidth]{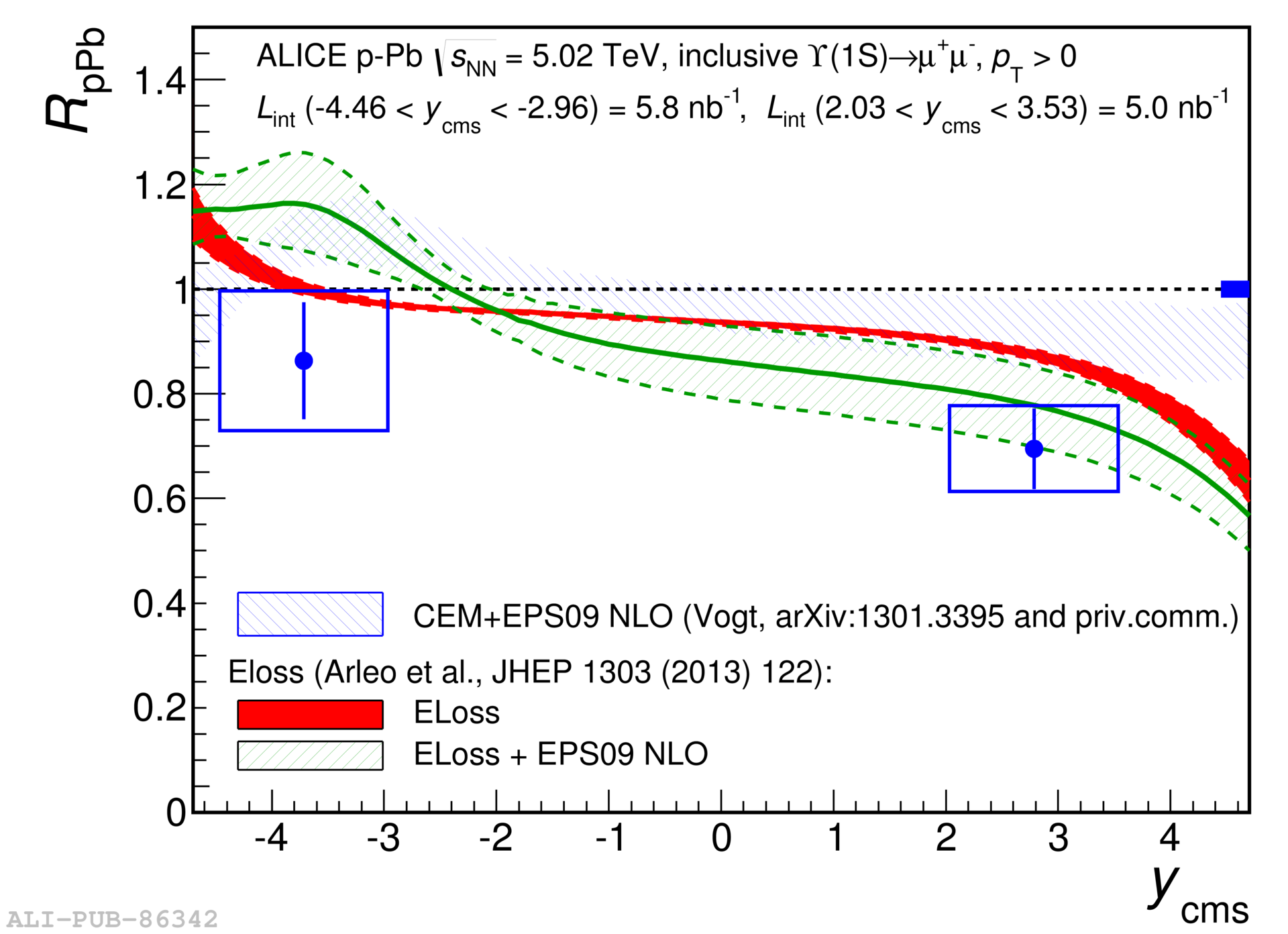}
\caption{The nuclear modification factor for inclusive $\Upsilon \rm (1S)$ production in p--Pb collisions at $\sqrt{s_{\rm{NN}}}$ = 5.02 TeV as a function 
of rapidity. Results are compared to those for J/$\psi$ (Left) as well as to different model predictions (Right). 
The vertical error bars represent the statistical uncertainties and the open boxes the uncorrelated systematic uncertainties. 
The filled boxes around $R_{\rm pPb} = 1$ show the size of the correlated uncertainties.}
\label{fig:figure3}
\end{figure}

\noindent The production of bottomonium in p--Pb systems was also studied by ALICE \cite{Abelev:2014oea}.
In Figure \ref{fig:figure3}, the measurements of the nuclear modification factor for the $\Upsilon \rm (1S)$ state as a function of rapidity 
in p--Pb collisions at $\sqrt{s_{\rm{NN}}}$ = 5.02 TeV are reported either differentially, in four rapidity intervals, or integrated over 
the backward or forward rapidity ranges. Data indicate a suppression of the inclusive $\Upsilon \rm (1S)$ production yields at forward 
rapidity in p--Pb compared to pp collisions, whereas at backward rapidity the $R_{\rm pPb}$ is compatible with unity within uncertainties.
When compared to previous J/$\psi$ measurements, the $\Upsilon \rm (1S)$ and J/$\psi$ $R_{\rm pPb}$ exhibit a similar trend, 
with the J/$\psi$ $R_{\rm pPb}$ being systematically higher at negative rapidities but rather similar to the 
$\Upsilon \rm (1S)$ $R_{\rm pPb}$ at positive rapidities within uncertainties. Results are compared to a model 
implementing a shadowing parametrization \cite{Albacete:2013ei}, and to a parton energy loss calculation \cite{Arleo:2012rs},
either with or without gluon shadowing.
% Despite the rather large experimental uncertainties, 
% none of the calculations is capable of fully describing the forward and backward rapidity data at the same time. 
% In particular, all calculations tend to overestimate the observed $R_{\rm pPb}$,
Even if the experimental uncertainties are large, the calculations tend to overestimate the measured nuclear modification factors, 
especially at backward rapidities, where a strong gluon anti-shadowing contribution appears disfavoured.
The $\Upsilon \rm (2S)$ production was also studied, and the ratio of the $\Upsilon \rm (2S)$ to $\Upsilon \rm (1S)$ 
production cross sections was measured in both rapidity ranges. Within uncertainties, compatibility with previous ALICE measurements 
of the same ratio in pp collisions at $\sqrt{s}$ = 7 TeV was found, therefore suggesting no evidence of a different magnitude of
CNM effects for the $\Upsilon \rm (2S)$ with respect to the $\Upsilon \rm (1S)$.  

\section{Conclusions}
 
Quarkonium production was measured in p--Pb collisions at $\sqrt{s_{\rm{NN}}}$ = 5.02 TeV by ALICE
as a function of $p_{\rm T}$, $y$ and centrality for different quarkonium states, and 
in p--Pb collisions at $\sqrt{s_{\rm{NN}}}$ = 8.16 TeV for the J/$\psi$ state. A significant suppression 
of the J/$\psi$ yield at forward rapidity and low $p_{\rm T}$ was observed, while the results appear compatible
at the two energies. The trend of the suppression can be fairly described by models 
implementing different CNM effects, such as shadowing and energy loss. 
A significantly stronger suppression of the $\psi \rm(2S)$ state relative to J/$\psi$ was measured in p--Pb collisions 
at $\sqrt{s_{\rm{NN}}}$ = 5.02 TeV. Models including final-state interactions are needed to explain the different degree of suppression. 
Measurements of $\Upsilon \rm (1S)$ production at $\sqrt{s_{\rm{NN}}}$ = 5.02 TeV show that the yield in p--Pb collisions is 
suppressed with respect to expectations from pp collisions at forward $y$, while at backward $y$ the data are consistent 
with no suppression within experimental uncertainties, disfavouring models with strong gluon anti-shadowing contributions.  

%%  if necessary
% \Acknowledgements
% I am grateful to XYZ for fruitful discussions.

\end{document}

%% file: econfmacros.tex
\def\beq{\begin{equation}}
\def\eeq#1{\label{#1}\end{equation}}
\def\eeqn{\end{equation}}

\def\beqa{\begin{eqnarray}}
\def\eeqa#1{\label{#1}\end{eqnarray}}
\def\eeqan{\end{eqnarray}}

\let\bar=\overbar

\def\Dslash{\not{\hbox{\kern-4pt $D$}}}
\def\dslash{\not{\hbox{\kern-2pt $\del$}}}

\def\msb{{\bar{\ssstyle M \kern -1pt S}}}